\documentclass[12pt]{article}
\usepackage{psfig}

\advance\textheight by\topskip

\begin{document}

\title{\bf
THE SOFTWARE SYSTEM ``EVOLUTION OF RADIO GALAXIES''
}

\author{
Oleg V. Verkhodanov$^1$,
\and Alexander I.Kopylov$^1$,
\and Olga P. Zhelenkova$^1$,
\and Natalia V. Verkhodanova$^1$,
\and Vladimir N. Chernenkov$^1$,
\and Yurij N. Parijskij$^1$,
\and {\it The Special Astrophysical Observatory of Russian Academy of Sciences}
\and{\it         357147 Karachai-Cherkesia, N.Arkhyz, Russia }
\and{\it E-mail: vo@sao.ru, akop@sao.ru, zhe@sao.ru, nat@sao.ru,
vch@sao.ru, par@sao.ru}
\and{\hspace*{20cm}}
\and Natalia S. Soboleva$^2$,
\and Adelina V. Temirova$^2$
\and {\it $^2$St.Petersburg branch of SAO, Pulkovo}
\and {\it E-mail: sns@fsao.spb.su, tem@fsao.spb.su}
}
\date{}
\maketitle

\begin{abstract}
The project of the informational system creation on the problem of
evolution of radio galaxies is described.
This system, being developed at present, will allow a user
to operate with simulated curves of spectral energy distributions (SED)
and to estimate ages and redshifts by photometrical data.
Authors use SEDs of several models (GISSEL96 (Bruzual, Charlot, 1996),
PEGASE (Fioc, Rocca-Volmerange, 1996) and Poggianti(1996))
for different types of galaxies.
Planned modes of access, formats of output result and additional functions
are described.
\end{abstract}

\section{Introduction}

The last few years have changed our view on the evolution of galaxies:
the most distant galaxies (i.e. systems having stellar population)
have been found at $z$=6.68 (Chen et al., 1999)
 and radio galaxies at $z$=5.2 (van Breugel et al., 1999).
Some of models suggests that galaxies can start their formation
at $z$=17 (Chen et al., 1999a). To understand deeper a situation
with a stellar population of galaxies and to check various mordern models
it is very important to have a capability to detect correctly
an age of galaxies.

The labour intensity of obtaining statistically significant high-quality data
on distant and faint galaxies and radio galaxies forces one to look for
simple indirect procedures in the determination of redshifts and other
characteristics of these objects. With regard to radio galaxies, even
photometric estimates turned out to be helpful and have so far been used
(McCarthy, 1993; Benn et al., 1989).

In the late 1980s and early 1990s it was shown that the color
characteristics
of galaxies can yield also the estimates of redshifts and ages
for the stellar
systems of the host galaxies. Numerous evolutionary models
appeared with which
observational data were compared to yield results strongly differed
from one another (Arimoto and Yoshii, 1987; Chambers and Charlot, 1990;
Lilly, 1987, 1990).

Over the last few years the three models: PEGASE
(Project de'Etude des Galaxies
par Synthese Evolutive (Fioc and Rocca-Volmerange, 1997)), Poggianti (1997)
and GISSEL96 (Bruzual, Charlot, 1996),
have been extensively used, in which an attempt has been made to eliminate the
shortcomings of the previous versions.

In the ``Big Trio'' experiment (Parijskij et al., 1996) we also attempted to
apply these techniques to distant objects of the RC catalogue with ultra steep
spectra (USS). Color data for nearly the whole basic sample of USS FR\,II
(Fanaroff and Riley, 1974) RC objects have been obtained with the 6\,m
telescope of SAO RAS.

To accelerate a procedure of age (and photometric redshift)
estimation we have begun a project ``Evolution of radio galaxies'',
supported by the Russian Foundation of Basic Research
(grant No.\,99-07-90334), which has to allow a user to obtain
age and photometric redshift estimations.

\section {Description of a system}

This system, being developed at present, will allow a user
to operate with simulated curves of spectral energy distributions (SED)
to estimate ages and redshifts by photometral data.
Authors use SEDs of three models:
\begin{itemize}
\item
    PEGASE (Fioc, Rocca-Volmerange, 1996) and
\item
    Poggianti(1996))
\item
    GISSEL96 (Bruzual, Charlot, 1996),
\end{itemize}
for different types of galaxies.

    The system will be situated on the special Web-server
unifying various type resources, including specialized Internet protocol
daemons (for the FTP, HTTP, e-mail support) and the designed
software permitting a user to operate with the SED curves.

    Requesting and filling in the standard
HTML-forms a user will be able to select different types of curves
or trust to do this to a computer by the $\chi^2$ method.
The input forms contain information about input filters or wavelengths
and corresponding magnitudes.

    There will be a possibility to detect the age of galaxies
in two ways:
\begin{enumerate}
\item
    fixed (known) redshift;
\item
    variable (unknown) redshift. In this case the redshift will be
    estimated too.
\end{enumerate}

The estimation of ages and redshifts iss performed by way of selection of the
optimum location on the SED curves
of the measured photometric points obtained when observing
radio galaxies in different filters. We use the already
computed table SED curves for different ages.
The algorithm of selection of the
optimum location of points on the curve consists briefly (for details see
Verkhodanov, 1996) in the following: by shifting the points lengthwise and
transverse the SED curve such a location was to be found at which the sum of
the squares of the discrepancies was a minimum. Through moving
over  wavelengths and flux density along the SED curve
we estimate the displacements
of the points from the location of the given filter and then the best fitted
positions were used to
compute the redshift. From the whole collection of curves, we select the
ones on which the sum of the squares of the discrepancies turned out to be
minimal for the given observations of radio galaxies.

There will be a chance to use infrared maps for absorption
estimates in a case when magnitudes are not corrected for absorption.

In order to take account of the absorption, we apply the maps (as
FITS-files)  from
the paper ``Maps of Dust IR Emission for Use in Estimation of Reddening and
CMBR Foregrounds'' (Schlegel et al., 1998).
 The conversion of stellar
magnitudes to flux densities are performed by the formula (e.g.
von Hoerner, 1974): $$S(Jy)=10^{C-0.4m}.$$

Another possibilities are supposed to be supported in our system:
\begin{itemize}
\item
    sorted bibliographical collection of papers for different stages of
       radio galaxy evolution,
\item
    archive of radio galaxies data in various
       wavelength ranges (both observed in Special astrophysical observatory
       and taken from Internet) and
\item
    search for information about radio galaxies
    using the largest data bases NED, CATS, LEDA et al.
    Very close interaction with the CATS database (Verkhodanov et al., 1996),
    designed and situated in the SAO, is supposed in the radio sources
    identification.
\end{itemize}

The HTTP and e-mail access to the basic procedures is organized
in this system.
Special e-mail forms are prepared to reflect input flags of low level
procedures.

FTP access is organized to give a user a possibility to obtain
SEDs models recorded in separate files.

A result is supposed to be written in ASCII tables and PS-figures and
can be sent to users.

An example of request execution, recorded in PS-files, is shown in Fig.1
(see detailed description and application of this system in
Verkhodanov et al. (1999)).

\begin{figure*}[!t]
\vbox{
\centerline{
\hspace*{-1.5cm}
\hbox{
\psfig{figure=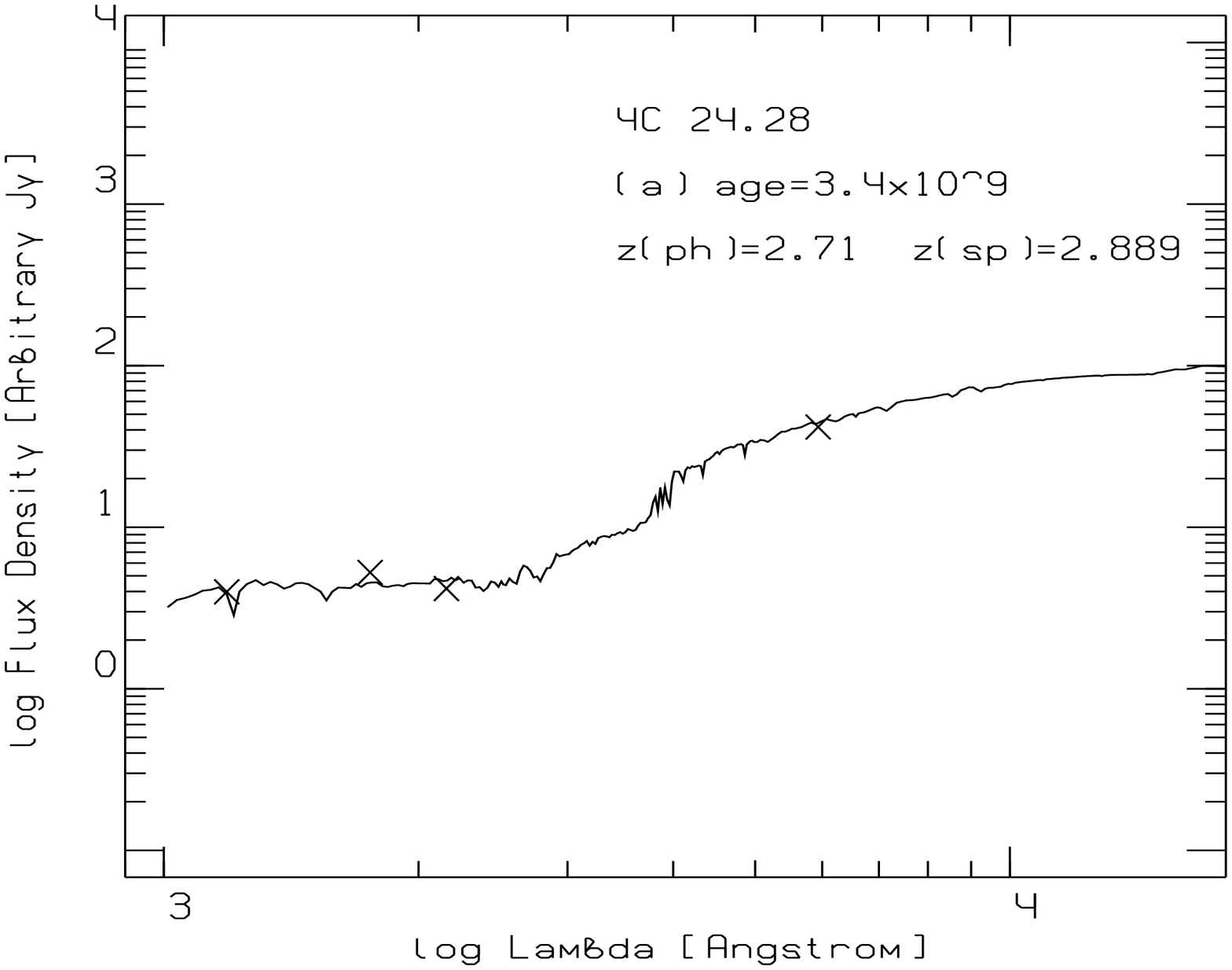,width=8cm,angle=-90}
\hspace*{-0.5cm}
\psfig{figure=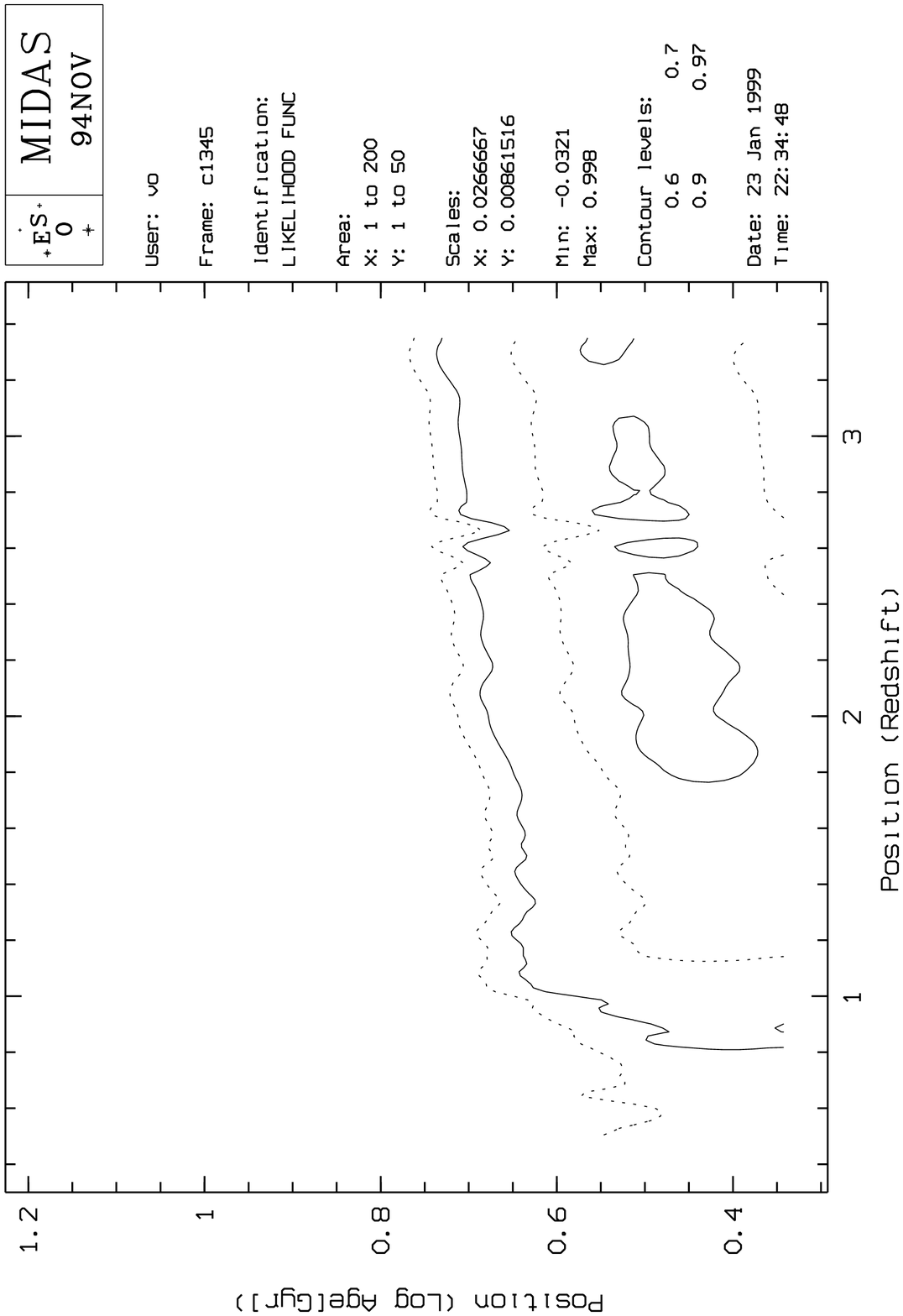,width=7cm,angle=-90,bbllx=43pt,bblly=58pt,bburx=571pt,bbury=660pt,clip=}
}}
\centerline{
\hspace*{-1.5cm}
\hbox{
\psfig{figure=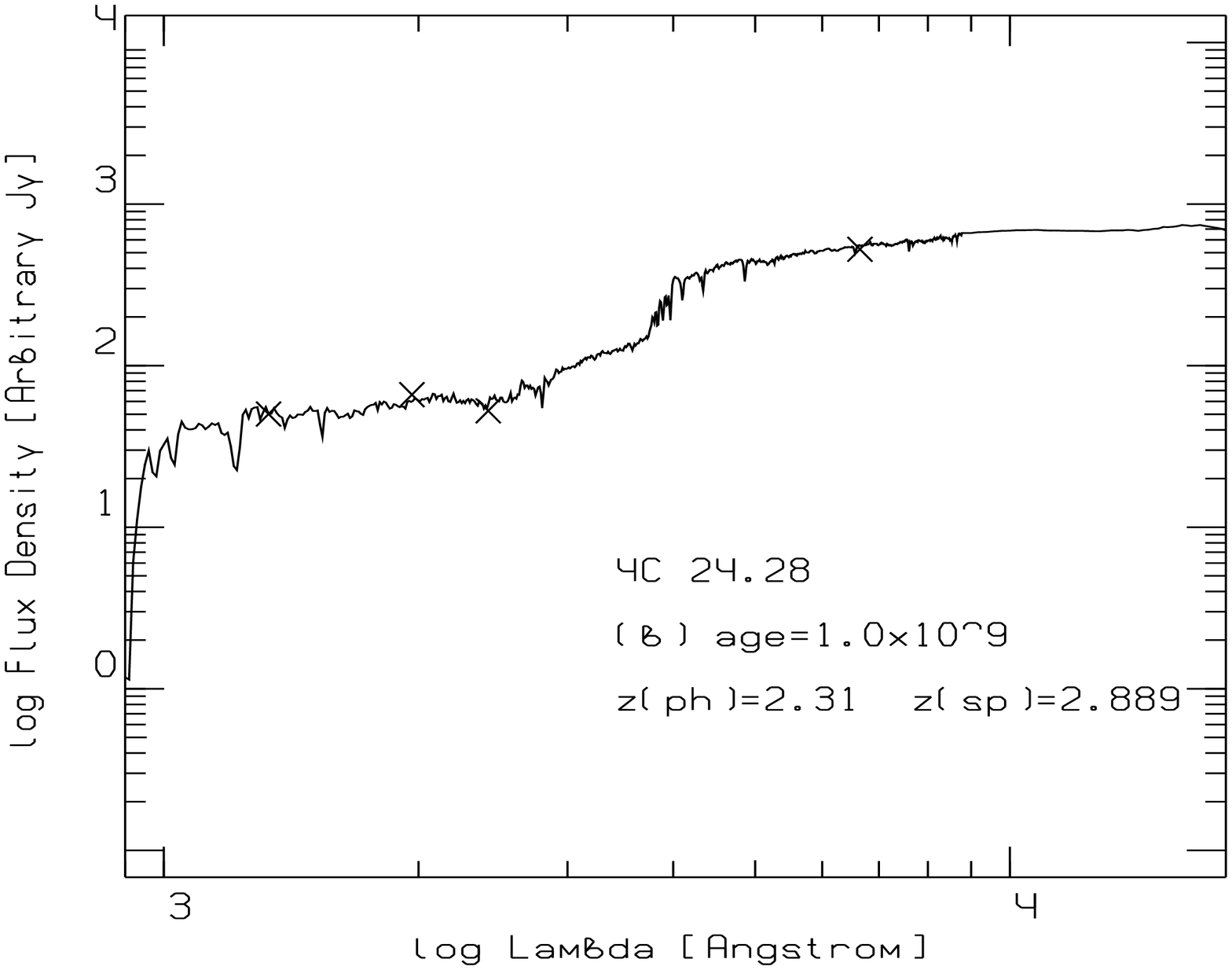,width=8cm,angle=-90}
\hspace*{-0.5cm}
\psfig{figure=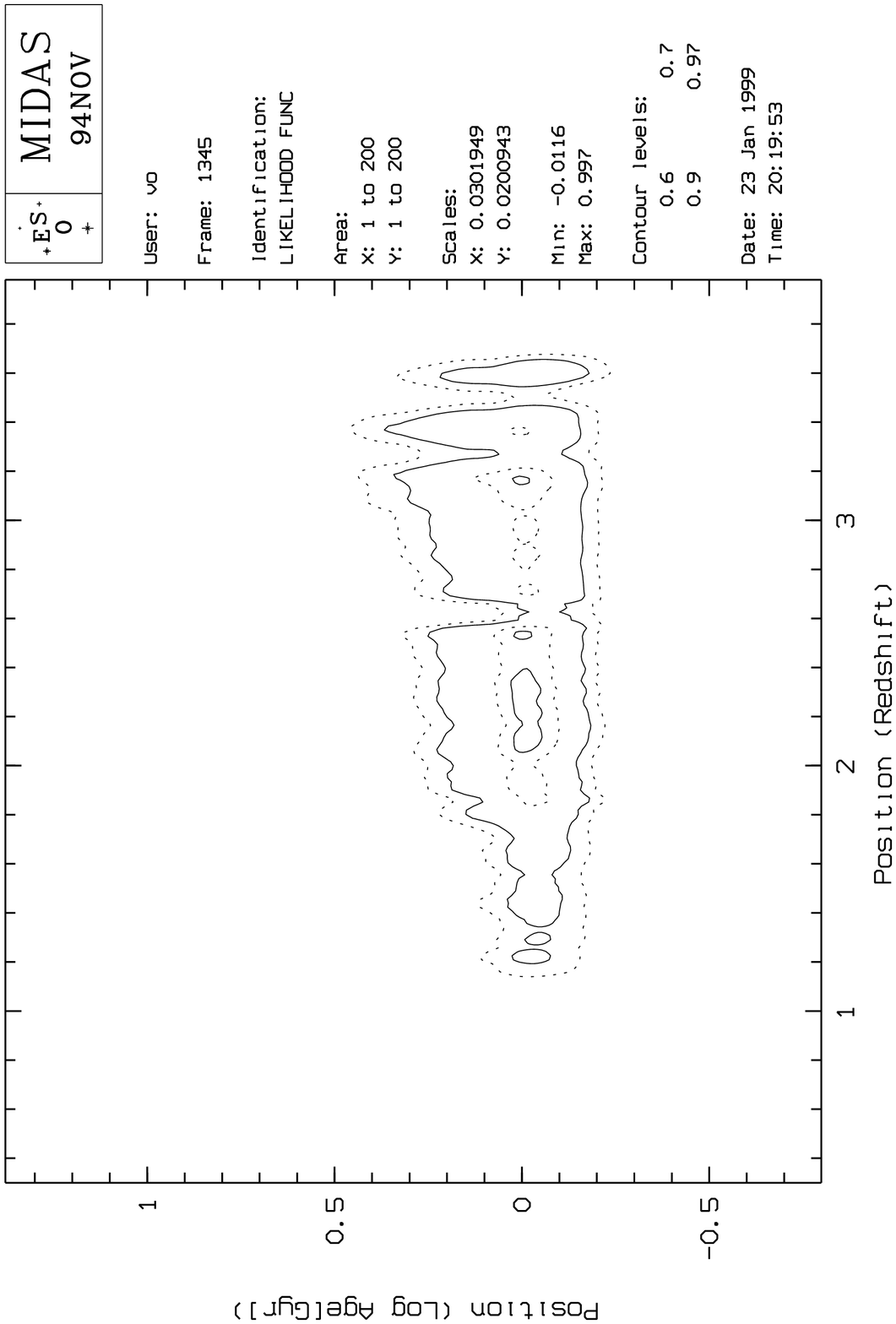,width=7cm,angle=-90,bbllx=43pt,bblly=58pt,bburx=571pt,bbury=660pt,clip=}
}}
\centerline{\parbox{16cm}{\caption{Estimation
of age and redshift for the 4C~24.28 radio galaxy
in the models of Poggianti
(a) and PEGASE (b).
SED curves with photometric data marked with crosses are shwon
on the left panels. Contours of likelihood functions vs redshift
and log~age are plotted at levels 0.6, 0.7, 0.9. 0.97 on the
right panels.
}}}}
\end{figure*}

\end{document}